\pdfoutput=1

\documentclass[a4paper,11pt]{article}

\usepackage{jheppub}

\usepackage[T1]{fontenc}
\usepackage[utf8]{inputenc}
\usepackage{amsmath,amssymb,amsfonts,mathtools}
\usepackage{bm}
\usepackage{slashed}
\usepackage{bbm}
\usepackage{graphicx}
\usepackage{subcaption}
\usepackage[compatibility=false]{caption}
\usepackage{booktabs}
\usepackage{array}
\usepackage{multirow}
\usepackage{tabularx}
\usepackage{makecell}
\usepackage{enumitem}
\usepackage{xcolor}
\usepackage{tikz}
\usepackage{pgfplots}
\pgfplotsset{compat=1.18}

\newcommand{\dd}{\mathrm{d}}
\newcommand{\ii}{\mathrm{i}}

\newcommand{\Imm}{\mathrm{Im}}

\title{Spectral functions in Minkowski quantum electrodynamics from neural reconstruction}

\author[a]{Rodrigo Carmo Terin}

\affiliation[a]{King Juan Carlos University, Faculty of Experimental Sciences and Technology,\\
Department of Applied Physics, Av. del Alcalde de Móstoles, 28933, Madrid, Spain}

\emailAdd{rodrigo.carmo@urjc.es}

\abstract{
We study neural reconstructions of quenched rainbow quantum electrodynamics (QED) Dyson--Schwinger benchmarks in Minkowski-related kinematics. 
Using the dispersive formulation as motivation, we separate the Euclidean Fukuda--Kugo equation, the spectral unitary equations, and the modified unitary equations. 
The Fukuda--Kugo benchmark is solved directly and shows the expected zero crossing above the critical region $\alpha_c=\pi/3$. 
Neural reconstructions with free output reproduce this behavior, while positivity-constrained ansätze fail in the supercritical regime. 
Thus, spectral positivity should be treated as a diagnostic of the Lehmann representation, not imposed blindly as a neural constraint.
}


\begin{document}

\maketitle

\flushbottom

\section{Introduction}
\label{sec:intro}

Dyson--Schwinger integral equations (DSEs) offer a continuum, nonperturbative
description of quantum field theory (QFT) through an infinite hierarchy of
coupled integral equations for Green functions.  Since the early works of
Dyson and Schwinger~\cite{Dyson1949,Schwinger1951,Schwinger1951b}, they have
become indispensable for studying dynamical mass generation, chiral symmetry
breaking, confinement and critical behavior in strongly interacting systems
\cite{Roberts1994,Alkofer:2000wg,Roberts2000,Fischer2019}.  Any practical
calculation requires truncating the hierarchy, and the reliability of the
result is tied to how well the truncation preserves the relevant symmetry,
analyticity and renormalization constraints.
QED is a natural testing ground for these
questions.  In the fermion sector, the propagator DSE captures the interplay
between renormalization, gauge covariance and dynamical mass generation.  The
simplest controlled starting point is the quenched rainbow approximation, in
which the fermion--photon vertex is replaced by its tree-level value and the
photon propagator is held fixed.  This truncation is not a complete
gauge-covariant description, but in the Landau gauge it provides a clean
benchmark for the scalar fermion dressing function. 

More sophisticated
constructions, such as the Ball--Chiu and Curtis--Pennington vertices
\cite{BallChiu1980,Curtis1990,Curtis1991}, implement Ward--Takahashi
identities and multiplicative renormalizability more accurately and are
essential for quantitative gauge-covariant studies.
A related Minkowski-space DSE formulation of QED was developed in
Ref.~\cite{OliveiraMacedoTerin2023}, where a minimal truncated set involving
the fermion and photon propagators, the photon--fermion vertex, and the
two-photon--two-fermion irreducible vertex was derived in general linear
covariant gauges, together with the corresponding Ward--Takahashi identities.
That construction shows both the relevance of symmetry-preserving
vertices and the complexity of going beyond the propagator-level truncation.
For the purposes of comparing Euclidean, Minkowskian and neural treatments of the same equation,
however, the rainbow--Landau setup remains a useful laboratory.

The central difficulty addressed here is not the existence of a DSE but the
metric in which it is solved.  Most continuum calculations are carried out in
Euclidean space, where numerical integration is more stable and singularity
structures are easier to handle.  Physical amplitudes, however, live in
Minkowski space, and the analytic continuation of Euclidean data is
frequently obstructed by branch cuts, thresholds, complex singularities and
pole motion
\cite{GimenoSegovia2008,PhysRevD.107.074026,PhysRevD.100.094001}.  Direct
Minkowski calculations therefore require a different approach, usually based
on spectral representations and subtracted dispersion relations.
The dispersive momentum-subtraction formulation ~\cite{Sauli2003JHEP}
is the central reference for the present work.  In that framework the
propagator is reconstructed from Lehmann weights, and renormalization is
implemented analytically through subtractions at a fixed momentum scale.  The
equations are written directly on the real axis, which makes it possible to
compare spacelike and timelike regimes within a single consistent treatment.
Fundamentally, the existence of a positive Lehmann representation is a physical
assumption in this formulation.  When the coupling becomes sufficiently
strong, the spectral solution may cease to exist or lose its particle
interpretation; this breakdown signals the absence of a fermion pole and is
commonly interpreted as a precursor to confinement-like behavior in
strong-coupling QED~\cite{FukudaKugo1976,Sauli2003JHEP}.
One must therefore be careful about which benchmark one is actually solving.
The Euclidean Fukuda--Kugo equation gives a spacelike benchmark for the
quenched rainbow gap equation and tracks the approach to the critical coupling
$\alpha_c = \pi/3$.  The system of unitary equations for the Lehmann weights
defines the genuinely spectral Minkowski construction.  And the modified
unitary equations introduce an infrared cutoff to regulate the threshold
enhancement of the spectral weights.  These three objects are related but not
interchangeable, and conflating them leads to confused comparisons.  In this
paper we keep them explicitly separate: the Fukuda--Kugo equation serves as
a controlled spacelike stress test, while the unitary and modified unitary
equations define the Minkowskian spectral target.

Physics-informed neural networks (PINNs) have emerged in recent years as
flexible solvers for differential and integral equations, representing the
residual of a physical equation directly in the loss function and turning the
network into a constrained function approximator~\cite{Lagaris1998,Raissi2019,
Karniadakis2021}.  PINN methods and their extensions have been applied to
fluid dynamics, cardiovascular flows, plasma physics, molecular modelling and
quantum systems~\cite{Raissi2020,Kissas2020,Mathews2020,Pfau2020,Zhang2018,
Yang2021b,Jagtap2020,Lu2021,Meng2020,Brevi2024,BreviTutorial2024}.  For DSEs,
the appeal is clear: a neural representation provides a differentiable
continuous approximation to propagator dressings and can incorporate nonlocal
residuals, renormalization constraints and asymptotic information within a
single optimization problem.
In our previous published work we applied PINNs to Euclidean QED DSEs by minimizing
integral-equation residuals for the fermion dressing functions
\cite{Terin2025SciPostCore}.  Multiscale losses and ultraviolet constraints
proved important for stabilizing the solution over many decades of momenta.
The present work extends that program to the Minkowskian setting, but with a
qualification that matters: constraints that stabilize a neural solver in weak
coupling can become physically misleading near the critical region.
Positivity and monotonicity work well as soft priors in subcritical regimes,
but they should not be imposed blindly when the benchmark itself develops zero
crossings or when the Lehmann representation is expected to fail.

Therefore our paper has two main goals.  The first is to formulate a Minkowskian neural
reconstruction of the scalar fermion dressing $B(p^2)$ under the same
quenched, rainbow, Landau-gauge assumptions used in the dispersive benchmark.
The neural loss combines residual terms, multiscale regularization and
separate spacelike/timelike stabilization.  The second is to stress-test this
framework using the Fukuda--Kugo equation.  Because this one is a
Volterra equation in $x = -p^2$, it can be solved directly without neural
approximation, providing a clean and independent reference.  We show that the
solution remains positive below the critical region, becomes strongly
suppressed near $\alpha_c = \pi/3$, and develops zero crossings above it.
An unconstrained neural ansatz reproduces this behavior faithfully, whereas a
positive-output ansatz fails in the supercritical regime by design: it cannot
represent a negative branch.
That failure is the main methodological point of our work.  Neural DSE solvers
should not be built solely to reproduce smooth weak-coupling solutions.  They
should also be capable of signaling when the basic assumptions are
breaking down.  Spectral positivity is best treated as a diagnostic of whether
a Lehmann representation remains valid, not as a fixed architectural
requirement.  This distinction matters especially for Minkowskian PINNs, where
analyticity, threshold behavior and the existence of physical poles are part
of the problem itself.

The paper is then organized as follows.  Section~\ref{sec:background} reviews the
Minkowskian spectral formulation, the fermion propagator decomposition,
subtracted dispersion relations and the role of Lehmann positivity.
Section~\ref{sec:methods} describes the dispersive benchmark and the neural
reconstruction strategy, with emphasis on the distinctions between the three
equations mentioned above.  Section~\ref{sec:results} presents the spacelike
and timelike benchmarks, the neural reconstructions, and the strong-coupling
stress test.  Section~\ref{sec:conclusions} summarizes the findings and
sketches natural extensions toward symmetry-preserving vertices, unquenched
dynamics, uncertainty-aware PINNs and bound-state equations.

\section{Background and spectral formulation}
\label{sec:background}

We work in Minkowski space-time with a metric convention in which timelike momenta
have $p^2 > 0$.  The renormalized fermion propagator is decomposed as
\begin{equation}
S^{-1}(p)
=
A(p^2)\slashed{p} - B(p^2),
\label{eq:fermion-prop-decomp}
\end{equation}
where $A(p^2)$ and $B(p^2)$ are the vector and scalar dressing functions.
The renormalization-group invariant mass function is
\begin{equation}
M(p^2) = \frac{B(p^2)}{A(p^2)}.
\label{eq:mass-function}
\end{equation}
In the quenched rainbow approximation the fermion--photon vertex is replaced
by its bare value,
\begin{equation}
\Gamma^\mu(p,k) = \gamma^\mu,
\end{equation}
and the photon propagator is held fixed.  In the Landau gauge this gives the
fermion DSE
\begin{equation}
S^{-1}(p)
=
Z_2\slashed{p} - Z_m m(\mu)
-
\ii e^2
\int \frac{\dd^4 k}{(2\pi)^4}\,
\gamma^\mu S(p-k)\gamma^\nu D_{\mu\nu}(k),
\label{eq:qed-dse-minkowski}
\end{equation}
with
\begin{equation}
D_{\mu\nu}(k)
=
\frac{-\ii}{k^2+\ii\epsilon}
\left(
g_{\mu\nu}
-
\frac{k_\mu k_\nu}{k^2}
\right)
\label{eq:landau-photon}
\end{equation}
in the massless quenched approximation.  This truncation is not a complete gauge-covariant treatment of QED, but a
controlled setting in which Euclidean, spectral Minkowskian and neural
reconstructions can be compared on equal footing.
In a general gauge-covariant truncation the fermion--photon vertex must satisfy
the Ward--Takahashi identity
\begin{equation}
(p-k)_\mu \Gamma^\mu(p,k)
=
S^{-1}(p) - S^{-1}(k).
\label{eq:WTI}
\end{equation}
The longitudinal part of the vertex is then fixed by the Ball--Chiu
construction~\cite{BallChiu1980}, and the equality $Z_1 = Z_2$ follows for
the vertex and wave-function renormalization constants.  The rainbow
calculation does not implement the full Ball--Chiu or Curtis--Pennington
structure, but in Landau gauge it gives a useful baseline for the scalar mass
equation and is the setup used in comparisons
discussed below.

The dispersive formulation rests on the assumption that, wherever particle-like
asymptotic states exist, the fermion propagator admits a Lehmann
representation.  Splitting a possible pole contribution from the continuum,
one writes
\begin{equation}
S(p)
=
\frac{r_f}{\slashed{p} - m + \ii\epsilon}
+
\int_{s_0}^{\infty}
\dd \omega\,
\frac{
\slashed{p}\,\sigma_v(\omega) + \sigma_s(\omega)
}{
p^2 - \omega + \ii\epsilon
},
\label{eq:fermion-lehmann}
\end{equation}
where $m$ is the pole mass when such a pole exists, $r_f$ is its residue,
and $\sigma_v, \sigma_s$ are continuum Lehmann weights.  In a confining or
supercritical regime this representation may no longer hold, and that
failure is one of the central diagnostics in the analysis of strong-coupling QED.
The fermion self-energy decomposes as
\begin{equation}
\Sigma(p) = \slashed{p}\,a(p^2) + b(p^2),
\label{eq:self-energy-decomp}
\end{equation}
so that
\begin{equation}
A(p^2) = A(\mu^2) - a(\mu;p^2),
\qquad
B(p^2) = A(\mu^2)m(\mu) + b(\mu;p^2).
\label{eq:A-B-from-self-energy}
\end{equation}
One subtraction suffices for the fermion self-energy in a
momentum-subtraction scheme.  The renormalized scalar functions are then
\begin{align}
a(\mu;p^2)
&=
\int_{s_0}^{\infty}
\dd s\,
\frac{\rho_v(s)(p^2 - \mu^2)}
{(p^2 - s + \ii\epsilon)(s - \mu^2)},
\label{eq:disp-a}
\\
b(\mu;p^2)
&=
\int_{s_0}^{\infty}
\dd s\,
\frac{\rho_s(s)(p^2 - \mu^2)}
{(p^2 - s + \ii\epsilon)(s - \mu^2)},
\label{eq:disp-b}
\end{align}
in which $\rho_v$ and $\rho_s$ are the absorptive parts of the vector and
scalar self-energies,
\begin{equation}
\Imm\,a(p^2) = \pi\rho_v(p^2),
\qquad
\Imm\,b(p^2) = \pi\rho_s(p^2),
\label{eq:absorptive-parts}
\end{equation}
on the cut.  With the normalization $A(\mu^2) = 1$, the scalar dressing
is reconstructed as
\begin{equation}
B(p^2)
=
m(\mu)
+
\int_{s_0}^{\infty}
\dd s\,
\frac{\rho_s(s)(p^2 - \mu^2)}
{(p^2 - s + \ii\epsilon)(s - \mu^2)},
\label{eq:B-dispersion}
\end{equation}
which guarantees
\begin{equation}
B(\mu^2) = m(\mu)
\label{eq:MOM-condition}
\end{equation}
and keeps the ultraviolet behavior finite within the chosen scheme.
In the massless quenched, rainbow, Landau-gauge approximation the vector
dressing simplifies to
\begin{equation}
A(p^2) = 1,
\label{eq:A-one}
\end{equation}
so that $M(p^2) = B(p^2)$.  This simplification is used throughout our
benchmark calculations, and should not be read as a general statement about
QED: a dressed vertex, a nonzero gauge parameter, or unquenching effects all
make $A(p^2)$ dynamical.
A physical asymptotic particle requires the spectral weights to define a
positive measure in the relevant channel.  If the propagator develops complex
singularities, loses its real pole, or fails to admit such a positive
representation, the particle interpretation breaks down.  In quantum chromodynamics (QCD) this is
ordinarily associated with confinement.  In strong-coupling quenched QED,
Fukuda and Kugo argued that the fermion pole can disappear above a critical
coupling~\cite{FukudaKugo1976}.  The dispersive analysis gives a
complementary spectral version of the same phenomenon: above a coupling of
order one, the Lehmann-weight equations cease to provide a reliable positive
spectral solution~\cite{Sauli2003JHEP}.
This has a direct consequence for neural reconstructions.  If the goal is to
learn a propagator in a regime where the Lehmann representation holds,
spectral positivity can serve as a useful diagnostic and, in weak coupling, a
stabilizing soft prior.  Near the critical region, however, imposing positivity
as a hard architectural constraint can suppress exactly the signal one wants to
observe.  For this reason positivity is used below as a diagnostic, not as a
universal constraint.
Although the numerical benchmarks in this work are quenched, it is useful to
recall the structure of the photon two-point function.  Gauge covariance
requires the vacuum polarization tensor to be transverse,
\begin{equation}
q_\mu \Pi^{\mu\nu}(q) = 0,
\end{equation}
and therefore
\begin{equation}
\Pi^{\mu\nu}(q)
=
\left(
g^{\mu\nu}
-
\frac{q^\mu q^\nu}{q^2}
\right)
\Pi(q^2).
\label{eq:vacpol-transverse}
\end{equation}
In a momentum-subtraction scheme the scalar polarization admits a once
subtracted dispersion relation,
\begin{equation}
\Pi_R(q^2,\mu^2)
=
\int_{s_\gamma}^{\infty}
\dd s\,\rho_\gamma(s)
\left[
\frac{1}{q^2 - s + \ii\epsilon}
-
\frac{1}{\mu^2 - s + \ii\epsilon}
\right],
\label{eq:photon-dispersion}
\end{equation}
with the Landau-gauge propagator
\begin{equation}
D_{\mu\nu}(q)
=
\frac{-\ii}{q^2[1 - \Pi_R(q^2,\mu^2)]}
\left(
g_{\mu\nu}
-
\frac{q_\mu q_\nu}{q^2}
\right).
\label{eq:photon-renormalized}
\end{equation}
In the quenched approximation, $\Pi_R$ is set to zero.  The unquenched
extension is conceptually straightforward but numerically more demanding,
since the fermion and photon spectral functions must then be solved
self-consistently.
For comparison with the spectral Minkowskian formulation, one also considers
the Euclidean gap equation.  In the Landau-gauge rainbow approximation, after
Wick rotation and angular integration, the scalar mass function satisfies
\begin{equation}
B(\zeta,x)
=
m(\zeta)
+
\frac{3\alpha}{4\pi}
\int_0^\infty
\dd y\,
V(\zeta,x,y)
\frac{B(\zeta,y)}
{y + B^2(\zeta,y)},
\label{eq:euclidean-gap-general}
\end{equation}
where $x = p_E^2 = -p^2$, $\zeta$ is the Euclidean subtraction point and
\begin{equation}
V(\zeta,x,y) = K(x,y) - K(\zeta,y).
\label{eq:euclidean-kernel-subtracted}
\end{equation}
Setting $\zeta = 0$ and $m(0) = 1$ gives the Fukuda--Kugo equation
\begin{equation}
B(0,x)
=
1 +
\frac{3\alpha}{4\pi}
\int_0^x
\dd y\,
\left(
\frac{y}{x} - 1
\right)
\frac{B(0,y)}
{y + B^2(0,y)}.
\label{eq:FKE-background}
\end{equation}
This equation is used below as a controlled spacelike stress test.  It is
not the Minkowskian spectral solution, but it exposes the approach to the
critical region and enables one to test, in a clean setting, whether a neural
representation can reproduce zero crossings when they appear.

\section{Methods}
\label{sec:methods}

This section describes the three numerical ingredients used in our paper: the
spectral/dispersive construction based on unitary equations, the direct
Euclidean Fukuda--Kugo benchmark, and the neural reconstruction strategy for
the scalar mass function $B(p^2)$.
The dispersive Minkowskian construction starts from the Lehmann
representation~\eqref{eq:fermion-lehmann}.  Inserting this representation
into the quenched rainbow DSE and taking absorptive parts yields a closed
system of real equations for the Lehmann weights.  In the massless-photon,
Landau-gauge approximation one has $A(p^2)=1$, so the dynamics is carried by
the scalar self-energy.  For a massless photon, the absorptive part of the
scalar self-energy is
\begin{equation}
\rho_s(\omega)
=
-3\left(\frac{e}{4\pi}\right)^2
\left[
r_f m
\left(
1-\frac{m^2}{\omega}
\right)
+
\int_{m^2}^{\omega}
\dd \alpha\,
\sigma_s(\alpha)
\left(
1-\frac{\alpha}{\omega}
\right)
\right],
\label{eq:rho-s-sauli}
\end{equation}
where $r_f$ is the pole residue, $m$ is the pole mass when it exists, and
$\sigma_s$ is the scalar Lehmann weight.  The Lehmann weights are obtained
from the unitary equations
\begin{align}
\sigma_v(\omega)
&=
\frac{f_1(\omega)+m(\mu)f_2(\omega)}
{\omega-m^2(\mu)},
\label{eq:sigma-v-ue}
\\
\sigma_s(\omega)
&=
\frac{m(\mu)f_1(\omega)+\omega f_2(\omega)}
{\omega-m^2(\mu)}.
\label{eq:sigma-s-ue}
\end{align}
In the simplified massless-photon case,
\begin{align}
f_1(\omega)
&=
r_f
\frac{m\rho_s(\omega)}{\omega-m^2}
+
[\sigma_s\ast\rho_s](\omega),
\label{eq:f1-ue}
\\
f_2(\omega)
&=
r_f
\frac{\rho_s(\omega)}{\omega-m^2}
+
[\sigma_v\ast\rho_s](\omega),
\label{eq:f2-ue}
\end{align}
with the principal-value convolution
\begin{equation}
[\sigma\ast\rho](\omega)
=
\mathcal{P}
\int_{m^2}^{\infty}
\dd x\,
\frac{
\rho(\omega)\sigma(x)\dfrac{x-\mu^2}{\omega-\mu^2}
+
\sigma(\omega)\rho(x)
}
{\omega-x}.
\label{eq:pv-convolution}
\end{equation}
The numerical difficulty of the spectral method is concentrated in these
principal-value terms, especially near threshold, where the Lehmann weights
can become strongly enhanced.
To regulate this infrared enhancement, the work \cite{Sauli2003JHEP} introduced modified unitary
equations in which the convolution terms are multiplied by
\begin{equation}
\Theta\!\left[\omega-m^2(1+c)\right],
\qquad
c=\frac{\alpha}{2\pi}.
\label{eq:mue-cutoff}
\end{equation}
Setting $c=0$ recovers the original unitary equations.  The modified
equations should therefore be regarded as an infrared-regulated numerical
variant of the spectral construction, not as a distinct field-theoretic
truncation.
Once $\rho_s$ is known, the scalar dressing is reconstructed from the
subtracted dispersion relation
\begin{equation}
B(p^2)
=
m(\mu)
+
\int_{m^2}^{\infty}
\dd \omega\,
\frac{\rho_s(\omega)(p^2-\mu^2)}
{(p^2-\omega+\ii\epsilon)(\omega-\mu^2)}.
\label{eq:B-reconstruction-method}
\end{equation}
For spacelike momenta, $p^2<0$, the integral is real.  For timelike
momenta, $p^2>0$, the $i\epsilon$ prescription fixes the real and
absorptive parts across the cut.
The Euclidean benchmark used below is the Fukuda--Kugo equation.  With the
choice $\zeta=0$ and $m(0)=1$, it reads
\begin{equation}
B(x)
=
1+
\frac{3\alpha}{4\pi}
\int_0^x
\dd y\,
\left(
\frac{y}{x}-1
\right)
\frac{B(y)}
{y+B^2(y)},
\qquad x=-p^2>0.
\label{eq:FKE-method}
\end{equation}
This is a Volterra equation.  Defining
\begin{equation}
F(y)
=
\frac{B(y)}{y+B^2(y)}
\label{eq:F-def}
\end{equation}
and
\begin{equation}
I_0(x)
=
\int_0^x \dd y\,F(y),
\qquad
I_1(x)
=
\int_0^x \dd y\,yF(y),
\label{eq:I0-I1}
\end{equation}
Eq.~\eqref{eq:FKE-method} becomes
\begin{equation}
B(x)
=
1+
\frac{3\alpha}{4\pi}
\left[
\frac{I_1(x)}{x}-I_0(x)
\right].
\label{eq:FKE-cumulative}
\end{equation}
Since the upper-endpoint contribution of the kernel vanishes, the solution can
be advanced sequentially on an ordered logarithmic grid in $x$.  We evaluate
$I_0$ and $I_1$ by cumulative trapezoidal quadrature.  The
Fukuda--Kugo solver is used as a controlled spacelike benchmark.  It is not
identified with the Minkowskian spectral solution.  Its role is to expose the
onset of the strong-coupling regime and to test whether neural ansätze can
reproduce zero crossings when they occur.
Our neural reconstruction is formulated for the spacelike variable
\begin{equation}
x=-p^2>0,
\qquad
t=\log_{10}x .
\end{equation}
We use a scalar feed-forward ansatz
\begin{equation}
N_\theta:\mathbb{R}\rightarrow\mathbb{R},
\qquad
t\mapsto N_\theta(t).
\end{equation}
Two output parametrizations are compared.  The unconstrained model is
\begin{equation}
B_\theta(x)
=
N_\theta(\log_{10}x),
\label{eq:Btheta-free}
\end{equation}
which can represent both positive and negative values.  The positive-output
model is
\begin{equation}
B_\theta^{+}(x)
=
\mathrm{softplus}
\left[
N_\theta(\log_{10}x)
\right]
+\delta,
\qquad
\delta>0,
\label{eq:Btheta-positive}
\end{equation}
which imposes $B_\theta^{+}(x)>0$. 
Also, $\theta$ denotes collectively all trainable weights and biases of the
network, and $t=\log_{10}x$ is the logarithmic input coordinate.  The
softplus function is defined as
$\mathrm{softplus}(z)=\log(1+e^z)$, so that
$B_\theta^{+}(x)$ is strictly positive.  The small constant $\delta>0$ is a
positive floor introduced only to avoid an exactly vanishing output.  The
brackets $\langle\cdots\rangle_x$ used below denote an average over the
chosen collocation points on the spacelike $x$ grid. The second model is used only as an
ablation test, because the Fukuda--Kugo benchmark becomes negative above the
critical region.
The physics-informed residual is obtained by inserting $B_\theta$ into the
Fukuda--Kugo equation:
\begin{equation}
\mathcal{R}_{\rm FKE}[B_\theta](x)
=
B_\theta(x)
-
1
-
\frac{3\alpha}{4\pi}
\left[
\frac{I_{1,\theta}(x)}{x}
-
I_{0,\theta}(x)
\right],
\label{eq:nn-fke-residual}
\end{equation}
where
\begin{equation}
I_{0,\theta}(x)
=
\int_0^x \dd y\,
\frac{B_\theta(y)}
{y+B_\theta^2(y)},
\qquad
I_{1,\theta}(x)
=
\int_0^x \dd y\,
y
\frac{B_\theta(y)}
{y+B_\theta^2(y)}.
\label{eq:neural-I0-I1}
\end{equation}
The corresponding physics loss is
\begin{equation}
\mathcal{L}_{\rm phys}
=
\left\langle
\left|
\mathcal{R}_{\rm FKE}[B_\theta](x)
\right|^2
\right\rangle_x.
\label{eq:nn-phys-loss}
\end{equation}
When the direct Fukuda--Kugo solution $B_{\rm FKE}$ is used as a reference,
we also include
\begin{equation}
\mathcal{L}_{\rm data}
=
\left\langle
\left|
B_\theta(x)-B_{\rm FKE}(x)
\right|^2
\right\rangle_x.
\label{eq:nn-data-loss}
\end{equation}
A mild smoothness penalty suppresses spurious high-frequency oscillations on
the logarithmic grid:
\begin{equation}
\mathcal{L}_{\rm sm}
=
\left\langle
\left|
\Delta_{\log}B_\theta(x)
\right|^2
\right\rangle_x.
\label{eq:nn-smooth-loss}
\end{equation}
Here $\Delta_{\log}$ denotes a finite-difference operator on the logarithmic
grid in $t=\log_{10}x$.  This term is not a physical constraint; it is only a
mild regularizer used to suppress grid-scale oscillations in the neural
output. The total loss is
\begin{equation}
\mathcal{L}_{\rm NN}
=
w_{\rm phys}\mathcal{L}_{\rm phys}
+
w_{\rm data}\mathcal{L}_{\rm data}
+
w_{\rm sm}\mathcal{L}_{\rm sm}.
\label{eq:nn-total-loss}
\end{equation}
The coefficients $w_{\rm phys}$, $w_{\rm data}$ and $w_{\rm sm}$ are
non-negative weights controlling, respectively, the physics residual, the
reference-data mismatch and the smoothness penalty. The relative reconstruction 
error is reported as
\begin{equation}
\varepsilon_2
=
\frac{
\|B_\theta-B_{\rm FKE}\|_2
}{
\|B_{\rm FKE}\|_2
}.
\label{eq:nn-relative-error}
\end{equation}
We also record the minimum value of $B_\theta$, the maximum absolute error
and the residual norm.  The comparison between
Eqs.~\eqref{eq:Btheta-free} and~\eqref{eq:Btheta-positive} isolates the
effect of imposing positivity on the neural output.
All integrations are performed on logarithmic grids.  For the Fukuda--Kugo
equation we use grids in $x=-p^2$ spanning several decades from the infrared
to the ultraviolet.  The zero-crossing diagnostic is obtained by recording
\begin{equation}
\min_x B(x)
\label{eq:minB}
\end{equation}
and the first value of $x$ at which $B(x)$ changes sign.  The
strong-coupling scan covers both subcritical and supercritical values of
$\alpha$, with particular attention to
\begin{equation}
\alpha_c=\frac{\pi}{3}.
\label{eq:critical-alpha}
\end{equation}
It is important to distinguish the analytical critical coupling from the finite-window diagnostic used in the numerical scan.  The former characterizes the onset of the critical asymptotic behavior
of the Fukuda--Kugo equation.  The latter records whether a zero crossing of
$B(x)$ is observed inside a finite spacelike interval $x\leq x_{\max}$.
Close to $\alpha_c$, the first zero crossing can be pushed to very large
values of $x$.  Therefore, a finite ultraviolet endpoint may lead to an
apparent zero-crossing threshold slightly above $\pi/3$, even when the
underlying critical value is unchanged.  In the refined scan reported below,
we vary $x_{\max}$ in order to separate this finite-window effect from a
genuine shift of the critical coupling.

The present neural reconstruction should be distinguished from a full spectral
PINN.  A full Minkowskian spectral PINN would learn both the scalar dressing
and the spectral density,
\begin{equation}
B_\theta(p^2),
\qquad
\rho_{s,\theta}(\omega),
\end{equation}
and would impose the subtracted dispersion relation as a residual.  Here we
use the Fukuda--Kugo equation as a controlled spacelike stress test.  The goal
is to determine whether neural constraints preserve or artificially remove the
strong-coupling features of a known DSE benchmark.

\section{Results}
\label{sec:results}

The discussion proceeds in three steps.  We first solve the Fukuda--Kugo
equation directly and use it as a spacelike benchmark for the approach to the
supercritical regime.  We then test neural reconstructions of this benchmark
and compare unconstrained and positive-output ansätze.  Finally, we connect
these results to the Minkowskian spectral problem and to unitary and
modified unitary equations.

\subsection{Direct Fukuda--Kugo benchmark}
\label{subsec:fke-results}

\begin{figure}
\centering
\includegraphics[width=0.72\linewidth]{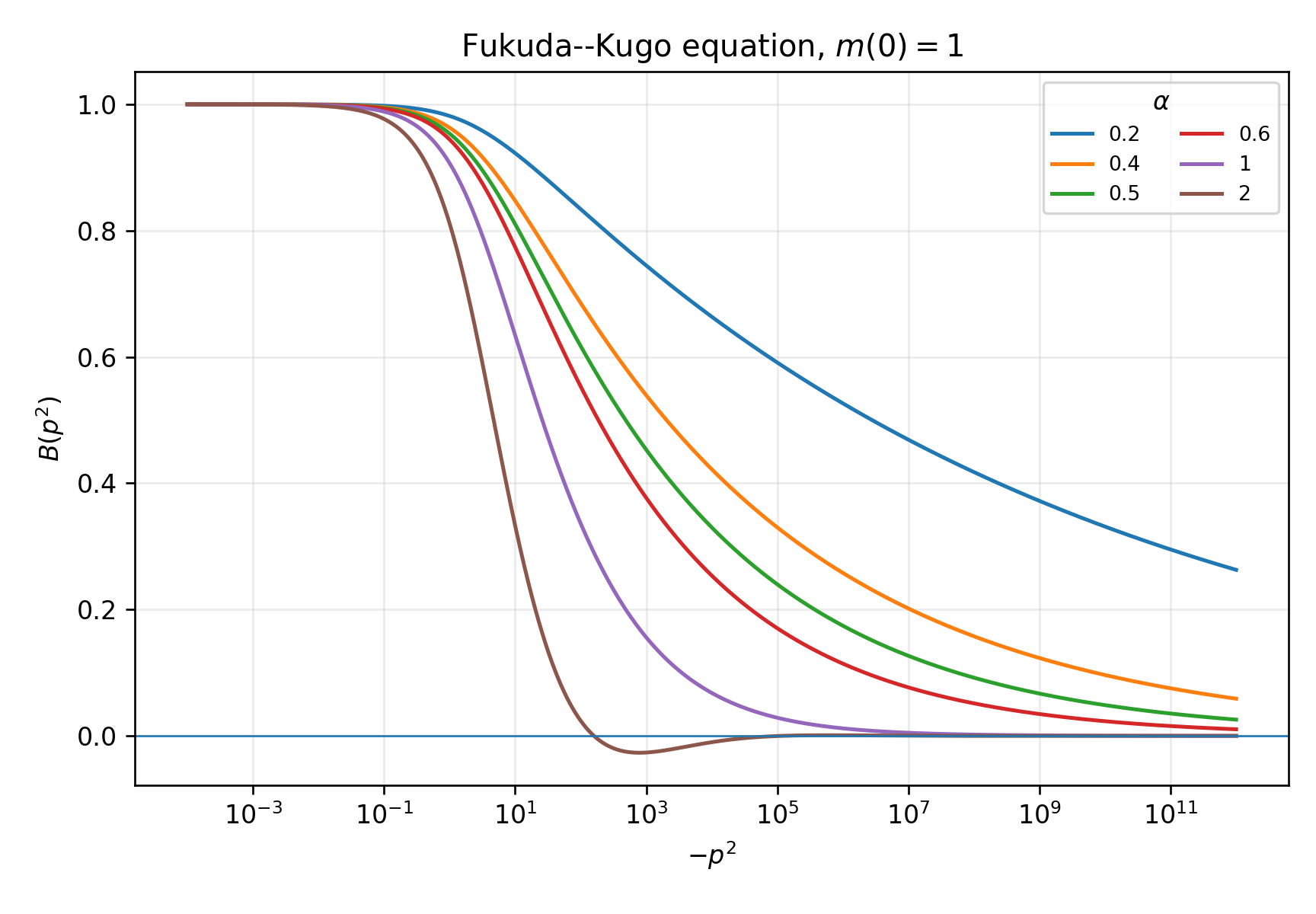}
\caption{ Direct solution of the Fukuda--Kugo equation for the normalization $m(0)=1$. The plotted quantity is the scalar mass function $B(x)$ as a function of the spacelike Euclidean variable $x=-p^2$, shown on a logarithmic horizontal scale. Different curves correspond to different values of the coupling $\alpha$, as indicated in the legend. For subcritical couplings the solution remains positive and decreases smoothly toward the ultraviolet, whereas for sufficiently large couplings the solution develops a negative branch. }
\label{fig:FKE-spacelike}
\end{figure}

\begin{figure}
\centering
\includegraphics[width=0.66\linewidth]{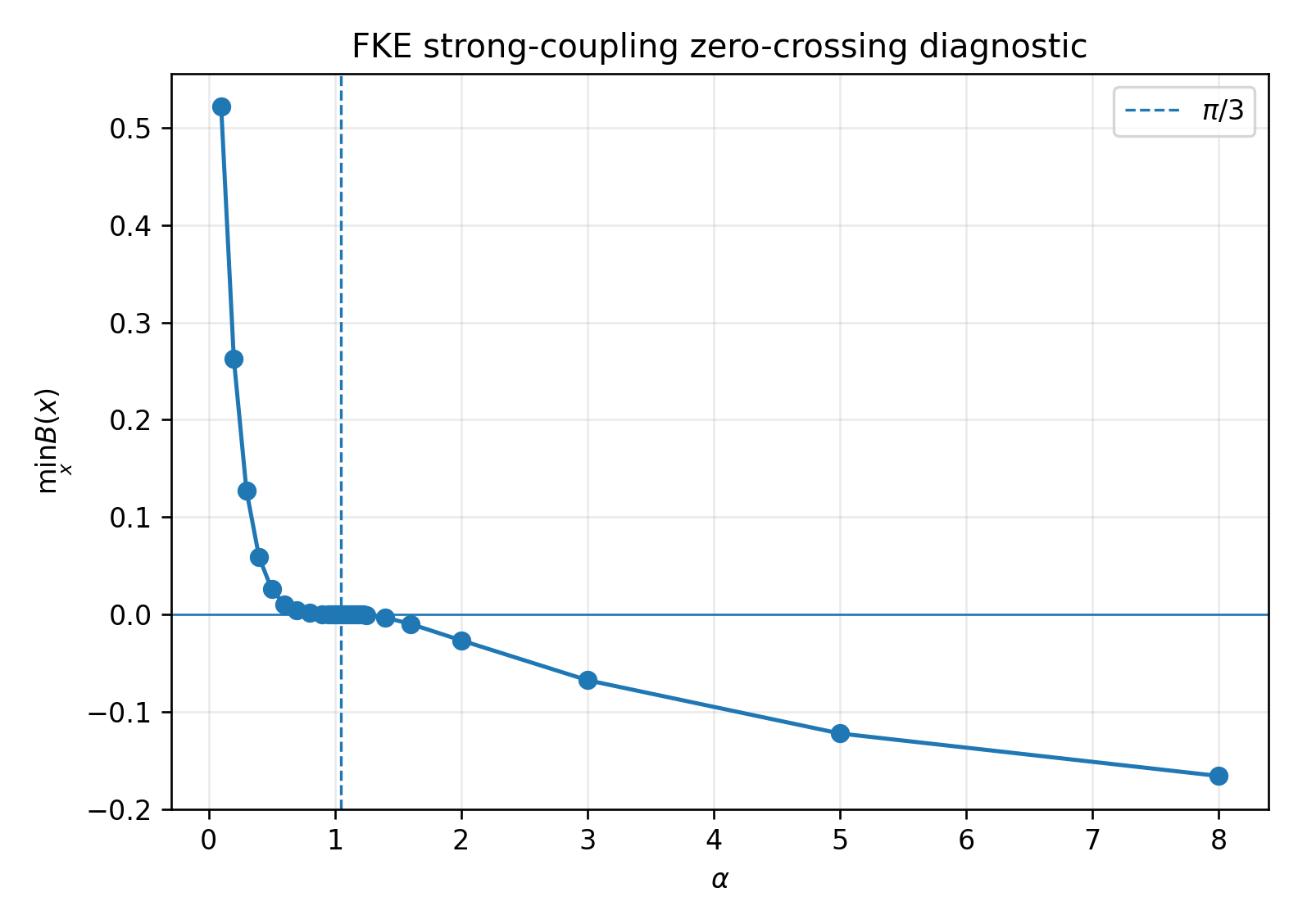}
\caption{
Strong-coupling zero-crossing diagnostic for the direct Fukuda--Kugo solution.
The plot shows the minimum value of the scalar mass function, $\min_x B(x)$,
over the numerical spacelike grid as a function of the coupling $\alpha$.
The horizontal line marks zero and the vertical dashed line indicates the
analytical critical value $\alpha_c=\pi/3$.  The finite-window diagnostic
records the first sign change observed within the chosen spacelike interval,
and is therefore sensitive to the ultraviolet endpoint and the resolution near
the critical region.
}
\label{fig:FKE-minB}
\end{figure}

Equation~\eqref{eq:FKE-method} is solved directly on a logarithmic grid in
$x = -p^2$ with the normalization $m(0) = 1$, using the cumulative
form~\eqref{eq:FKE-cumulative}.  This calculation involves no neural network
and provides a clean reference for the behavior of the mass function in the
spacelike domain.
Figure~\ref{fig:FKE-spacelike} shows the resulting mass function for several
values of the coupling.  At weak and intermediate couplings the solution is
positive and decreases smoothly toward the ultraviolet.  As $\alpha$
approaches the critical region, the ultraviolet tail becomes strongly
suppressed, and for couplings slightly above $\alpha_c = \pi/3$ the solution
develops a zero crossing and turns negative over part of the spacelike domain.
A more quantitative picture is given by the minimum value of the mass function
over the spacelike grid.  Figure~\ref{fig:FKE-minB} shows $\min_x B(x)$ as
a function of $\alpha$, with the vertical dashed line marking
$\alpha_c = \pi/3$.  The minimum stays positive below the critical region
and becomes negative shortly above it.
The zero-crossing scan is summarized in Table~\ref{tab:FKE-scan}.  On the
main numerical interval used in the calculation, $x_{\max}=10^{12}$, the
first observed zero crossing appears between $\alpha=1.075$ and
$\alpha=1.10$.  This should not be interpreted as an independent numerical
determination of the analytical critical coupling.  Rather, it is a
finite-window diagnostic: it records whether a sign change occurs within the
chosen spacelike interval.

\begin{figure}
\centering
\includegraphics[width=0.72\linewidth]{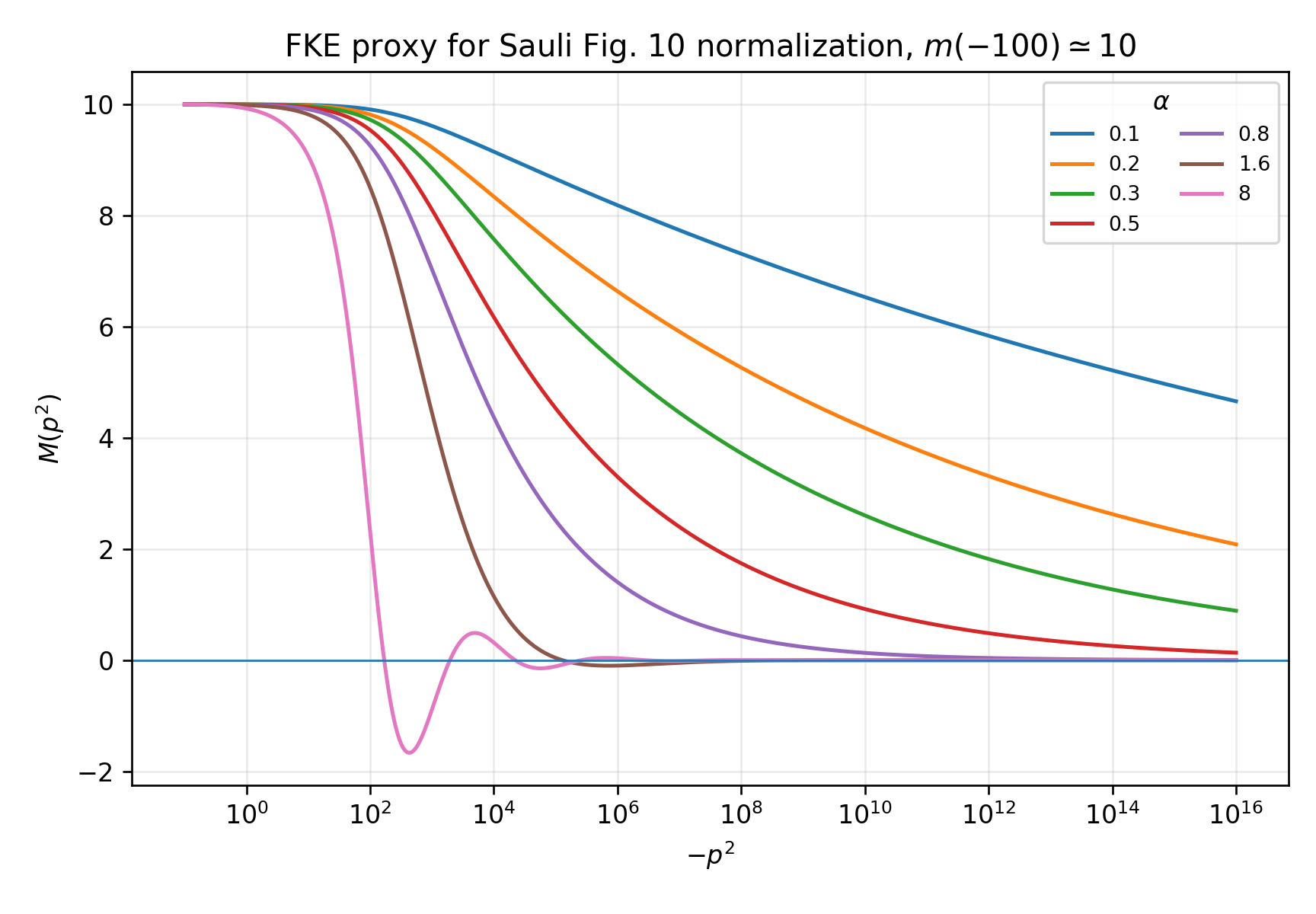}
\caption{ Scaled Fukuda--Kugo spacelike benchmark in a normalization chosen to resemble the standard dispersive-method comparison with $m(-100)\simeq 10$. The horizontal axis is the spacelike momentum variable $-p^2$, shown on a logarithmic scale, and the vertical axis shows the corresponding normalized mass function. The figure is used only as a shape and normalization check: it illustrates how increasing $\alpha$ suppresses the ultraviolet tail and, for large enough couplings, produces oscillatory sign-changing behavior. }
\label{fig:FKE-sauli10}
\end{figure}

\begin{table}
\centering
\begin{tabular}{c c c c}
\toprule
$\alpha$ & $\min_x B(x)$ & $B_{\rm UV}$ & zero crossing \\
\midrule
$0.8$   & $1.24\times 10^{-3}$  & $1.24\times 10^{-3}$  & no \\
$0.9$   & $3.22\times 10^{-4}$  & $3.22\times 10^{-4}$  & no \\
$1.0$   & $5.35\times 10^{-5}$  & $5.35\times 10^{-5}$  & no \\
$1.05$  & $1.37\times 10^{-5}$  & $1.37\times 10^{-5}$  & no \\
$1.075$ & $4.45\times 10^{-6}$  & $4.45\times 10^{-6}$  & no \\
$1.10$  & $-5.71\times 10^{-7}$ & $-5.71\times 10^{-7}$ & yes \\
$1.20$  & $-2.07\times 10^{-4}$ & $-1.82\times 10^{-6}$ & yes \\
$2.0$   & $-2.67\times 10^{-2}$ & $1.06\times 10^{-6}$  & yes \\
$8.0$   & $-1.66\times 10^{-1}$ & $1.32\times 10^{-7}$  & yes \\
\bottomrule
\end{tabular}
\caption{Strong-coupling scan of the direct Fukuda--Kugo equation on the main numerical
spacelike grid.  The column $\min_x B(x)$ gives the minimum value of the
mass function over the grid, $B_{\rm UV}$ denotes the value at the largest
spacelike momentum included in the grid, and the last column records whether
a sign change is observed within the finite numerical interval.  The
zero-crossing entry should therefore be interpreted as a finite-window
diagnostic, not as an independent determination of the analytical critical
coupling.
}
\label{tab:FKE-scan}
\end{table}
To clarify this point, we repeated the scan near $\alpha_c=\pi/3$ while
varying the ultraviolet endpoint $x_{\max}$.  The result is shown in
Table~\ref{tab:FKE-window}.  For $x_{\max}=10^{12}$, no zero crossing is
observed up to $\alpha=1.095$, while the first sign change appears at
$\alpha=1.10$, around $x\simeq 3.37\times 10^{11}$.  As $x_{\max}$ is
increased, the first observed finite-window threshold moves toward
$\pi/3$.  Thus the apparent mismatch between $\pi/3$ and the coarser
estimate $1.075$--$1.10$ is not evidence for a shifted critical coupling;
it reflects the fact that, close to criticality, the first zero crossing is
pushed to very large spacelike momenta.
\newpage
\begin{table}
\centering
\begingroup
\color{blue}
\begin{tabular}{c c c}
\toprule
$x_{\max}$ & first $\alpha$ with zero crossing & first zero $x_0$ \\
\midrule
$10^{12}$ & $1.100$ & $3.37\times 10^{11}$ \\
$10^{16}$ & $1.080$ & $6.22\times 10^{14}$ \\
$10^{20}$ & $1.070$ & $7.39\times 10^{17}$ \\
$10^{24}$ & $1.065$ & $2.02\times 10^{20}$ \\
\bottomrule
\end{tabular}
\endgroup
\caption{Dependence of the zero-crossing diagnostic on the ultraviolet endpoint
$x_{\max}$.  The second column gives the first value of $\alpha$, within
the refined scan, for which a zero crossing of the direct Fukuda--Kugo
solution is observed inside the finite interval $x\leq x_{\max}$.  The
third column gives the corresponding first zero $x_0$, obtained by linear
interpolation between adjacent grid points.  The drift of this finite-window
threshold toward $\pi/3$ as $x_{\max}$ is increased shows that the
apparent shift is a finite-domain effect, not a modification of the analytical
critical coupling.
}
\label{tab:FKE-window}
\end{table}
Figure~\ref{fig:FKE-sauli10} shows the corresponding scaled spacelike curves
in a normalization chosen to resemble dispersive method comparison at
$m(-100) \simeq 10$.  This figure is included only as a shape and
normalization check against the standard presentation of the strong-coupling
spacelike solutions; it does not add new physics.
\newpage
\subsection{Neural reconstruction of the Fukuda--Kugo benchmark}
\label{subsec:neural-fke-results}

\begin{figure}
\centering
\includegraphics[width=0.76\linewidth]{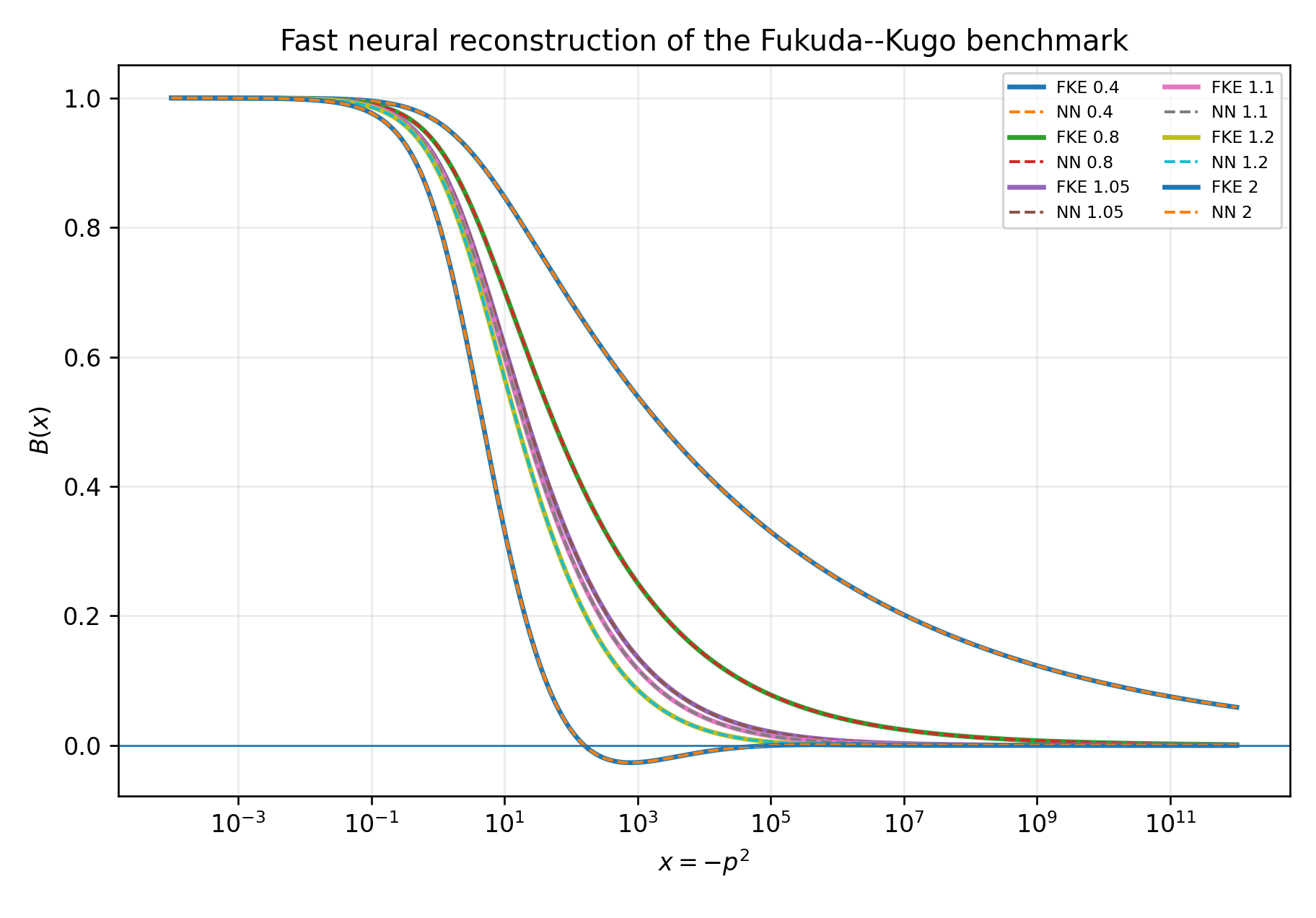}
\caption{ Comparison between the direct Fukuda--Kugo solution $B_{\rm FKE}(x)$ and the free-output neural reconstruction $B_\theta(x)$. The horizontal axis is the spacelike variable $x=-p^2$, shown on a logarithmic scale, and the vertical axis is $B(x)$. For each coupling $\alpha$, the solid curve shows the direct Fukuda--Kugo benchmark and the corresponding dashed curve shows the neural reconstruction. The close overlap demonstrates that the unconstrained neural ansatz can reproduce both positive subcritical solutions and sign-changing supercritical solutions.}
\label{fig:NN-FKE}
\end{figure}

\begin{figure}
\centering
\includegraphics[width=0.68\linewidth]{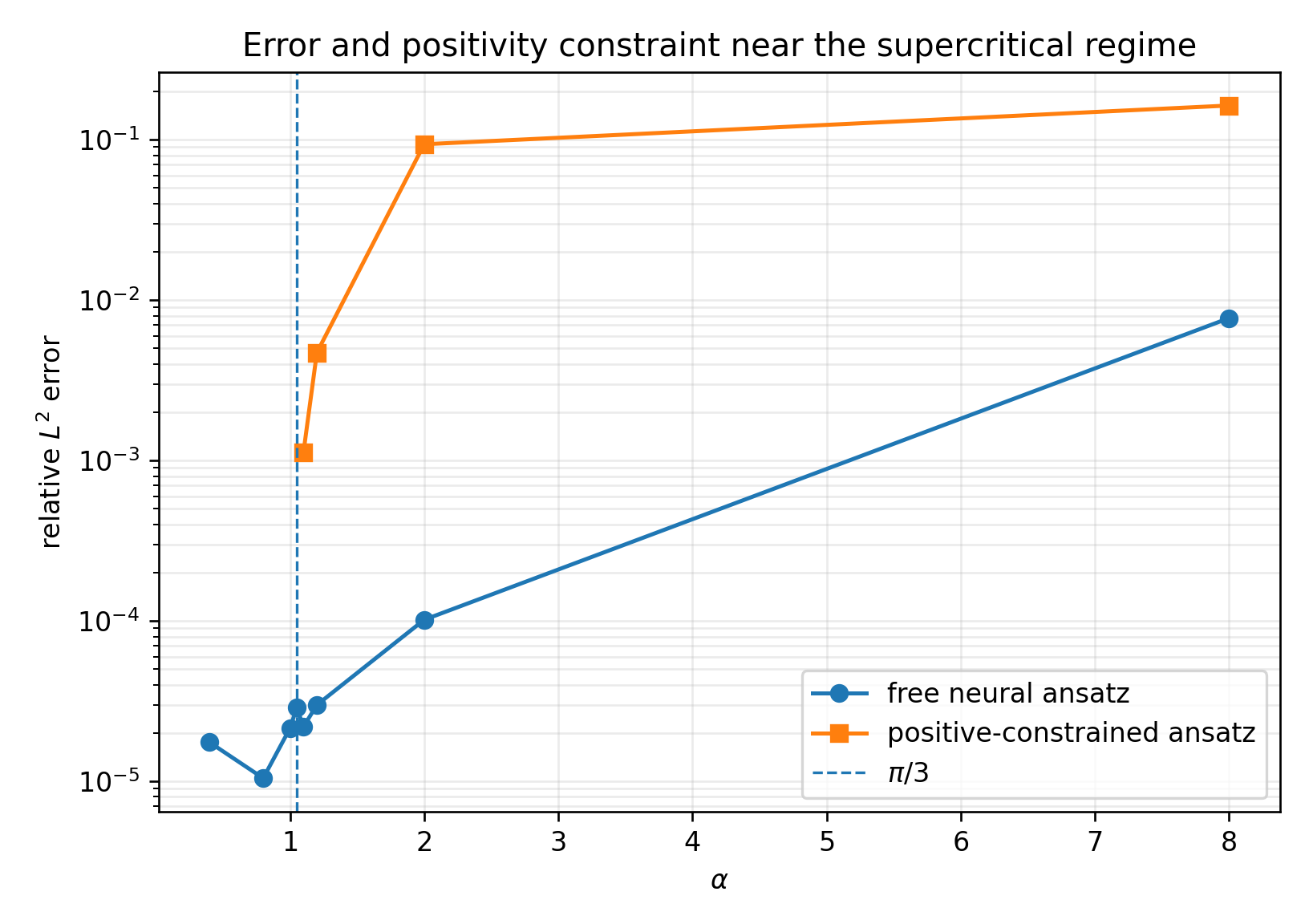}
\caption{ Relative $L^2$ reconstruction error as a function of the coupling $\alpha$. The blue circular markers correspond to the free-output neural ansatz, while the orange square markers correspond to the positive-output ansatz. The vertical dashed line marks $\alpha_c=\pi/3$. The error of the free ansatz remains small across the displayed range, whereas the positive-output ansatz deteriorates rapidly once the reference Fukuda--Kugo solution develops a negative branch. }
\label{fig:NN-error}
\end{figure}

We next use the direct Fukuda--Kugo solution as a benchmark for neural
reconstruction.  The purpose of this test is not to replace the dispersive
Minkowskian calculation, but to isolate a concrete problem in neural DSE loss
design: constraints that help in weak coupling may become incorrect near the
supercritical regime.
We compare the free-output and positive-output ansätze by measuring the
relative error and the residual of the Fukuda--Kugo equation.
Figure~\ref{fig:NN-FKE} compares the direct solution with the unconstrained
neural reconstruction.  The neural ansatz reproduces the solution accurately
from weak coupling through the near-critical region and into the supercritical
regime, including the zero crossing and the negative branch where these
appear.
The relative errors are shown in Fig.~\ref{fig:NN-error} and summarized in
Table~\ref{tab:NN-FKE-error}.  The unconstrained ansatz remains accurate
over the full range of couplings.  The positive-output ansatz, by contrast,
deteriorates faster once the benchmark turns negative.  This is not an
optimization failure; it is the result of an output constraint that conflicts
with the benchmark.

\begin{figure}
\centering
\includegraphics[width=0.68\linewidth]{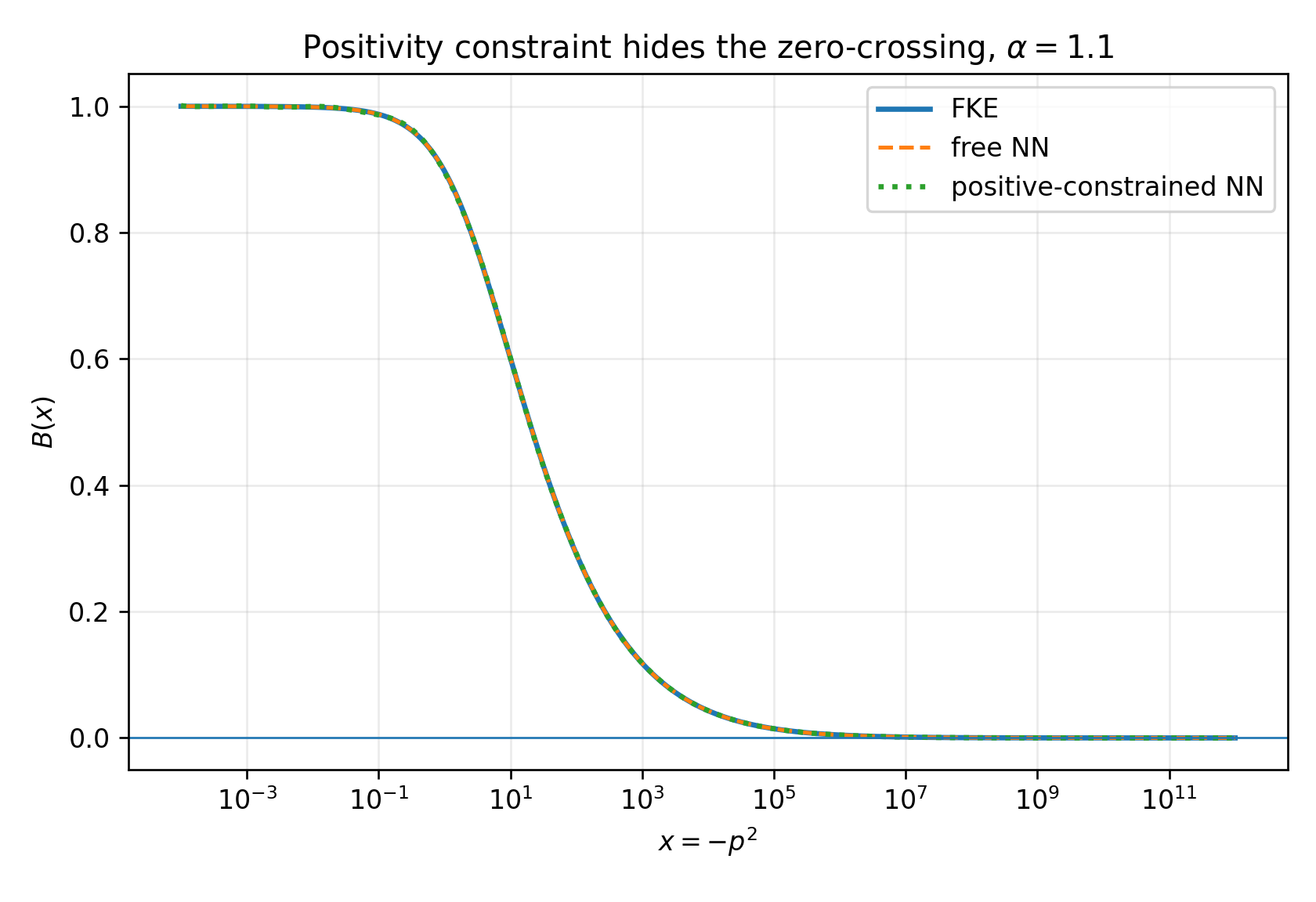}
\caption{ Effect of imposing positivity on the neural output at $\alpha=1.1$. The solid curve is the direct Fukuda--Kugo solution, the dashed curve is the free-output neural reconstruction, and the dotted curve is the positive-constrained reconstruction. At this coupling the negative branch is very small on the displayed vertical scale, but the positive-output ansatz is already unable to represent the correct sign structure. This case illustrates the onset of the failure of hard positivity constraints near the supercritical regime. }
\label{fig:positivity-alpha11}
\end{figure}

\begin{figure}
\centering
\includegraphics[width=0.68\linewidth]{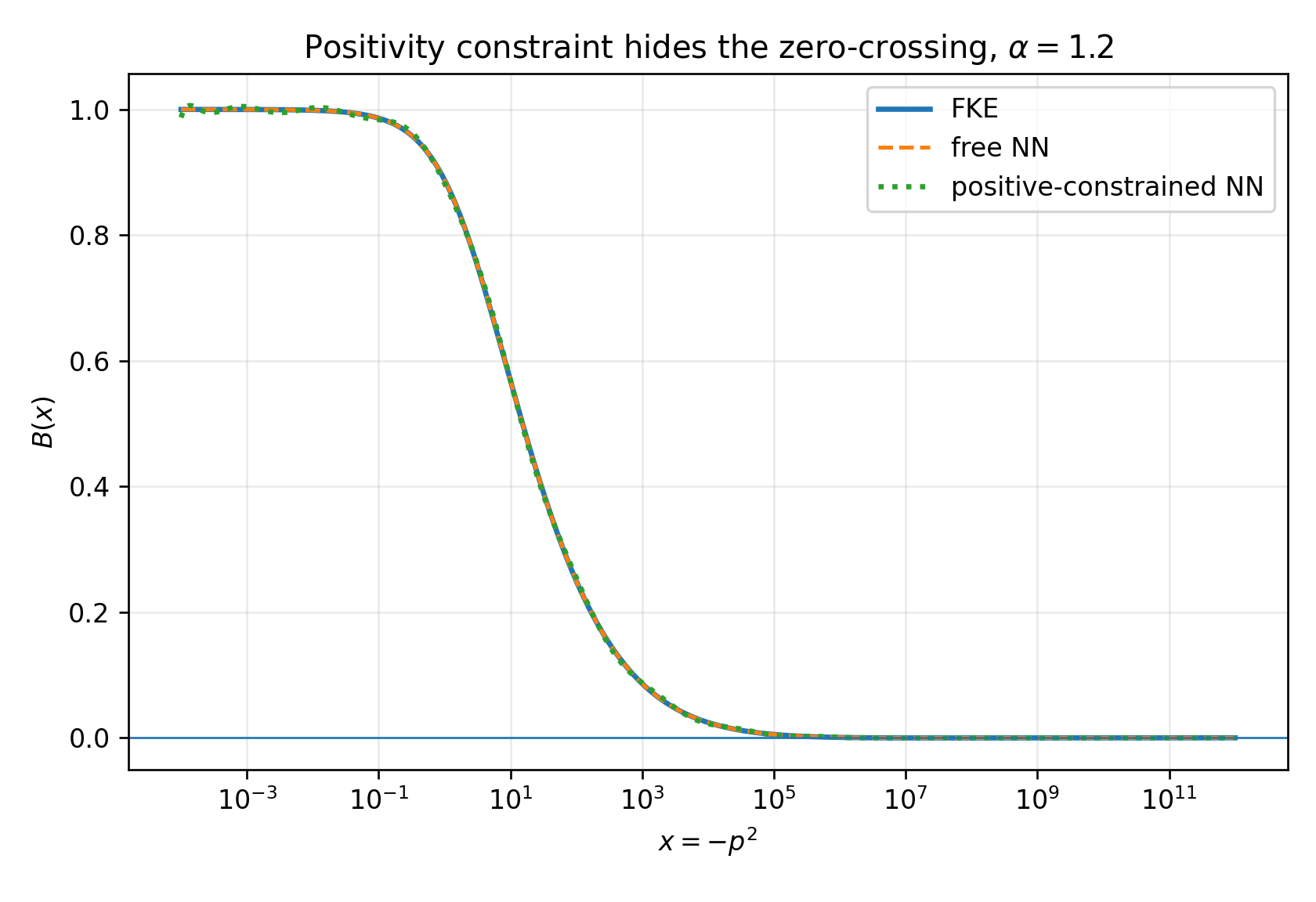}
\caption{ Effect of imposing positivity on the neural output at $\alpha=1.2$. The solid curve shows the direct Fukuda--Kugo benchmark, the dashed curve shows the free-output neural reconstruction, and the dotted curve shows the positive-constrained reconstruction. The free ansatz follows the reference solution, including its small negative branch, whereas the positive ansatz is forced to remain non-negative and therefore removes the sign-changing feature by construction. }
\label{fig:positivity-alpha12}
\end{figure}

\begin{figure}
\centering
\includegraphics[width=0.68\linewidth]{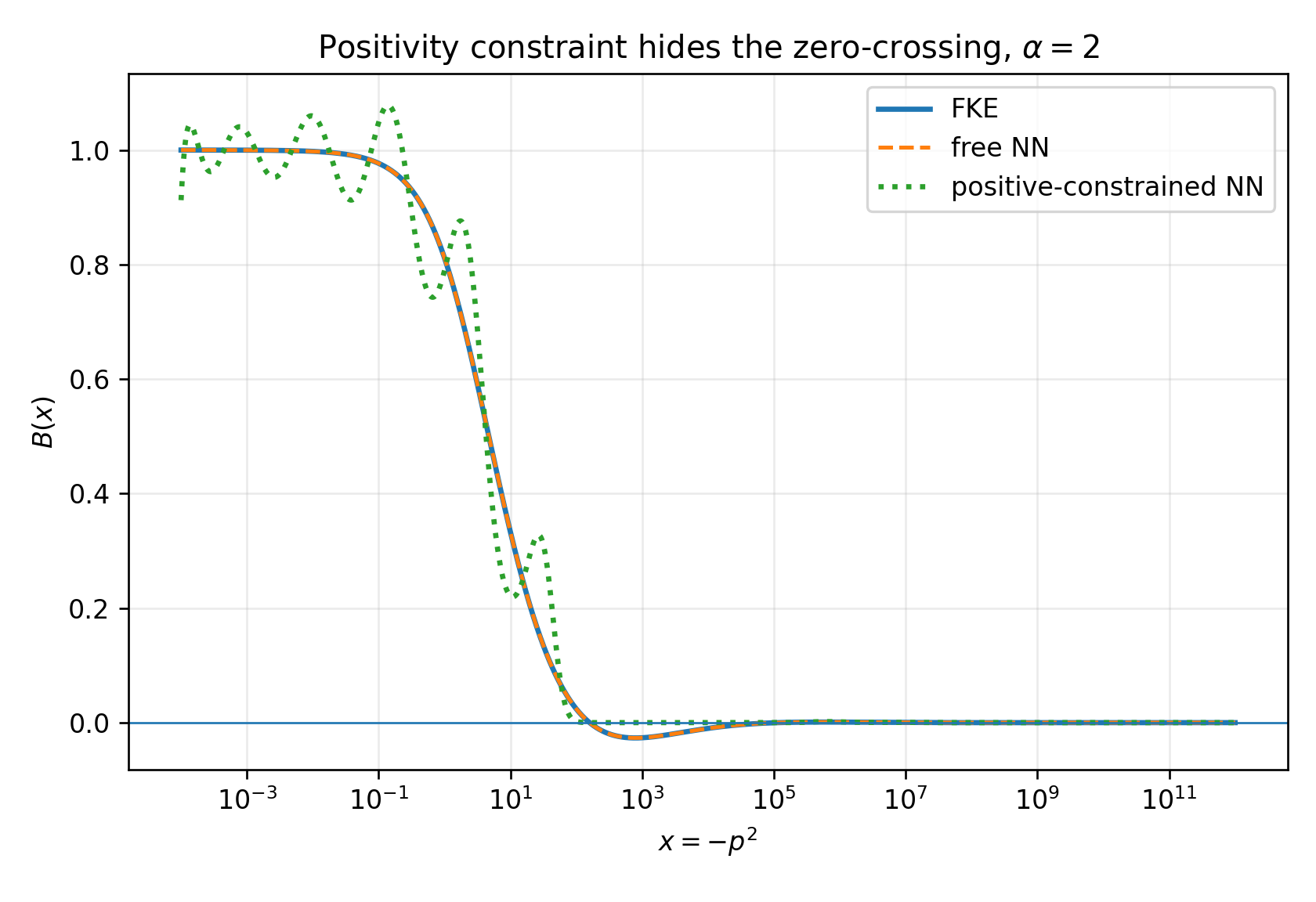}
\caption{ Effect of imposing positivity on the neural output at $\alpha=2.0$. The solid curve is the direct Fukuda--Kugo solution, the dashed curve is the free-output neural reconstruction, and the dotted curve is the positive-constrained reconstruction. The unconstrained network follows the negative branch of the benchmark, while the positive-output network cannot cross zero and instead produces a distorted non-negative approximation. }
\label{fig:positivity-alpha2}
\end{figure}

\begin{figure}
\centering
\includegraphics[width=0.68\linewidth]{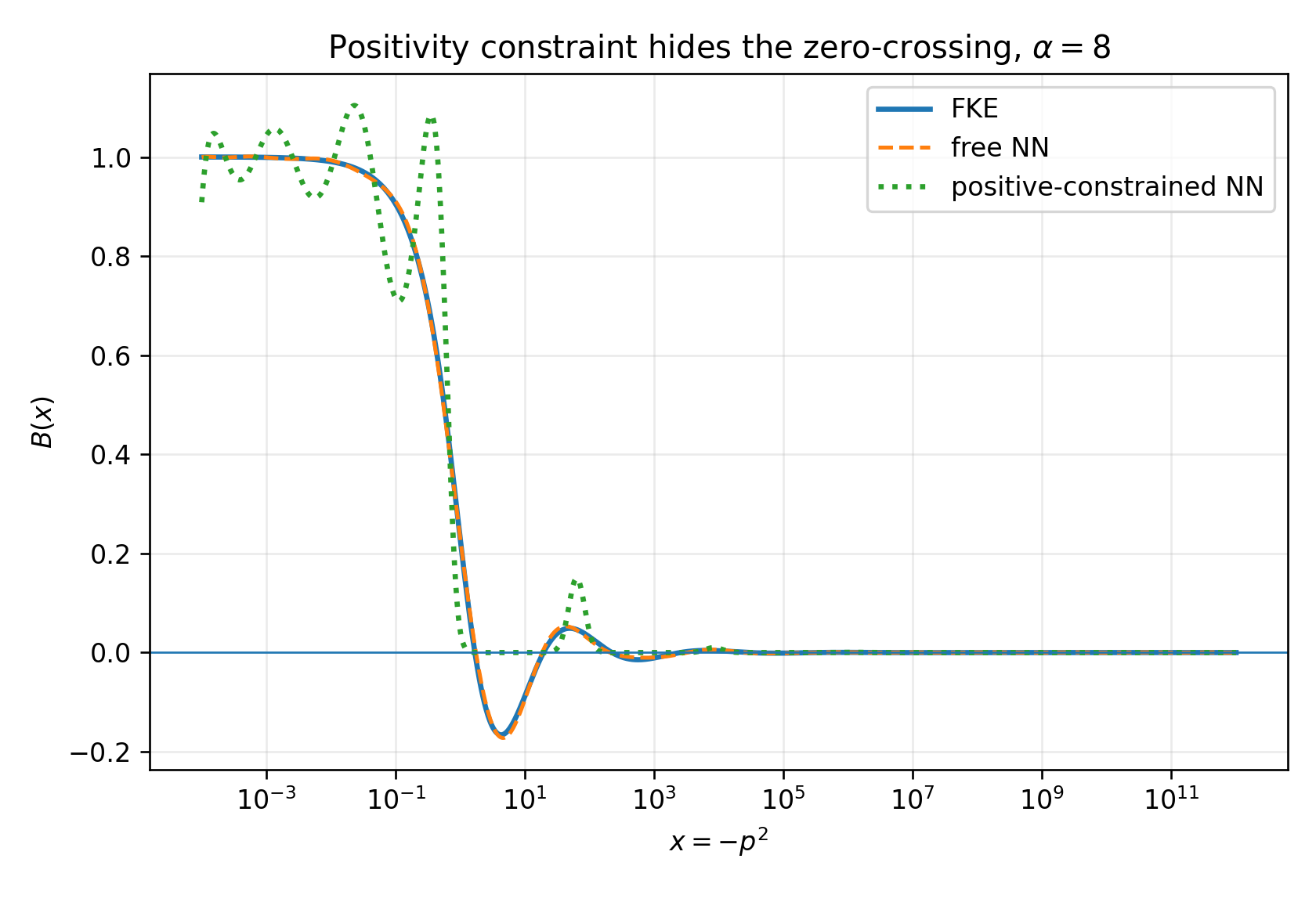}
\caption{ Effect of imposing positivity on the neural output at $\alpha=8.0$. The solid curve is the direct Fukuda--Kugo solution, the dashed curve is the free-output neural reconstruction, and the dotted curve is the positive-constrained reconstruction. In this strongly supercritical case the reference solution has a pronounced negative branch. The free ansatz tracks this behavior, whereas the positive-output ansatz hides the zero crossing and develops spurious positive oscillations. }
\label{fig:positivity-alpha8}
\end{figure}

\begin{table}
\centering
\begin{tabular}{c c c c c}
\toprule
$\alpha$ & model & $\varepsilon_2$ &
$\min B_{\rm FKE}$ & $\min B_\theta$ \\
\midrule
0.4 & free & $1.8\times 10^{-5}$ &
$5.89\times 10^{-2}$ & $5.89\times 10^{-2}$ \\
0.8 & free & $1.0\times 10^{-5}$ &
$1.24\times 10^{-3}$ & $1.23\times 10^{-3}$ \\
1.0 & free & $2.1\times 10^{-5}$ &
$5.34\times 10^{-5}$ & $4.65\times 10^{-5}$ \\
1.1 & free & $2.2\times 10^{-5}$ &
$-5.81\times 10^{-7}$ & $-1.41\times 10^{-5}$ \\
1.1 & positive & $1.1\times 10^{-3}$ &
$-5.81\times 10^{-7}$ & $1.75\times 10^{-7}$ \\
1.2 & free & $3.0\times 10^{-5}$ &
$-2.08\times 10^{-4}$ & $-2.07\times 10^{-4}$ \\
1.2 & positive & $4.7\times 10^{-3}$ &
$-2.08\times 10^{-4}$ & $1.16\times 10^{-7}$ \\
2.0 & free & $1.0\times 10^{-4}$ &
$-2.67\times 10^{-2}$ & $-2.67\times 10^{-2}$ \\
2.0 & positive & $9.4\times 10^{-2}$ &
$-2.67\times 10^{-2}$ & $7.18\times 10^{-8}$ \\
8.0 & free & $7.7\times 10^{-3}$ &
$-1.66\times 10^{-1}$ & $-1.72\times 10^{-1}$ \\
8.0 & positive & $1.6\times 10^{-1}$ &
$-1.66\times 10^{-1}$ & $2.43\times 10^{-8}$ \\
\bottomrule
\end{tabular}
\caption{ Neural reconstruction errors for the Fukuda--Kugo benchmark. The relative error $\varepsilon_2$ is defined in Eq.~\eqref{eq:nn-relative-error}. The columns $\min B_{\rm FKE}$ and $\min B_\theta$ compare the minimum value of the direct benchmark with that of the neural reconstruction. The free-output ansatz can reproduce negative branches, whereas the positive-output ansatz is constrained to remain non-negative and therefore fails once the benchmark changes sign. }
\label{tab:NN-FKE-error}
\end{table}
The ablation plots make this effect vivid.
Figure~\ref{fig:positivity-alpha11} shows the first supercritical case, where
the zero crossing is barely visible: even there, the positive-output model
cannot represent the correct sign structure.
Figures~\ref{fig:positivity-alpha12}--\ref{fig:positivity-alpha8} show that
the discrepancy grows steadily as the coupling increases.

\begin{figure}
\centering
\includegraphics[width=0.72\linewidth]{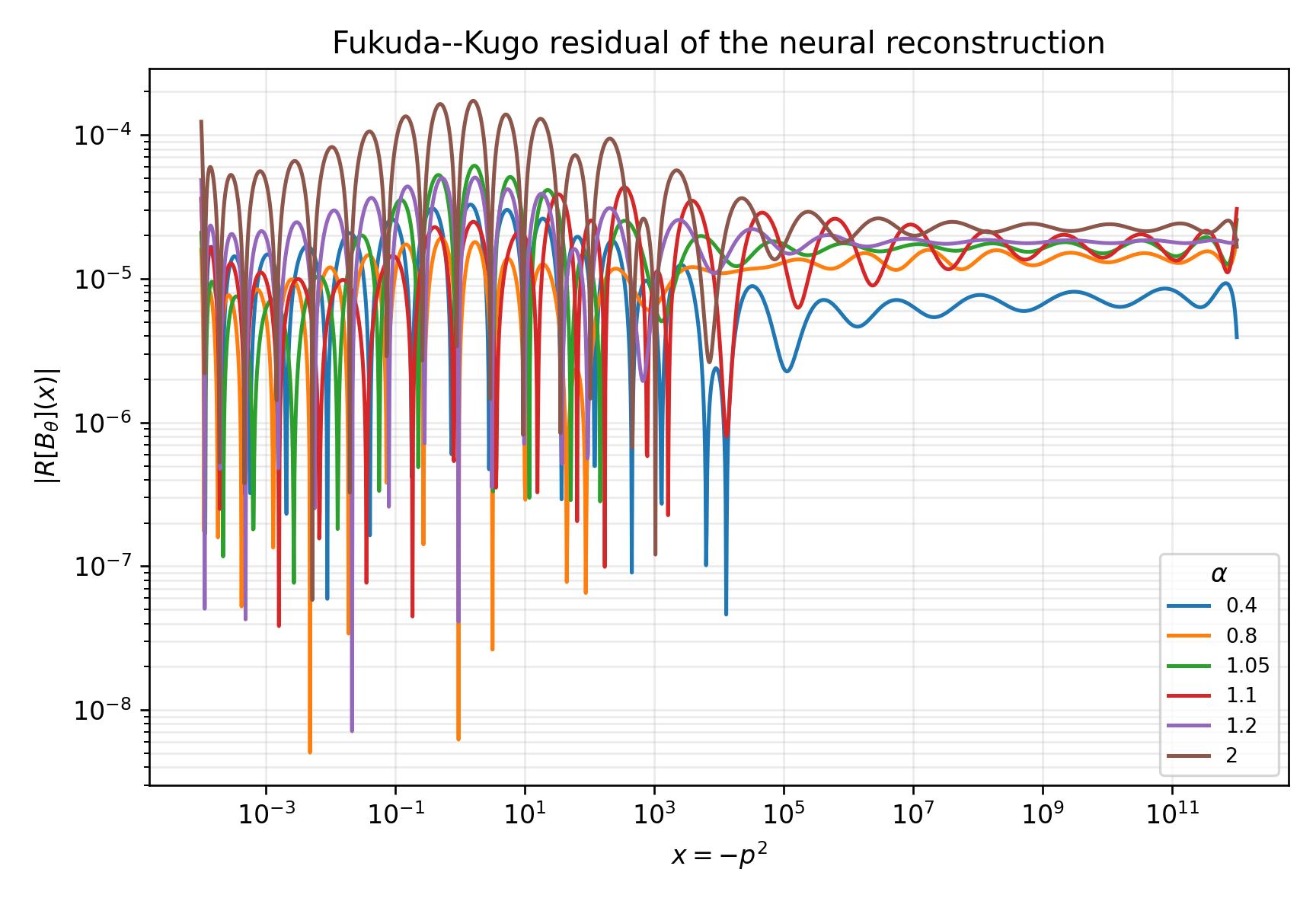}
\caption{ Absolute value of the Fukuda--Kugo residual evaluated on the free-output neural reconstructions. The residual is $|\mathcal{R}_{\rm FKE}[B_\theta](x)|$, plotted as a function of the spacelike variable $x=-p^2$ on logarithmic axes. Different curves correspond to different couplings $\alpha$. The residuals remain small over the full grid, with larger values appearing in regions where the mass function becomes strongly suppressed or changes rapidly. }
\label{fig:NN-residuals}
\end{figure}

The residuals of the unconstrained neural reconstructions are shown in
Fig.~\ref{fig:NN-residuals}.  They stay small over the grid, with the
largest deviations appearing near regions where the benchmark changes quickly.
Near the supercritical region the mass function becomes strongly suppressed
and sign changes amplify relative residuals, so some growth there is expected.

\subsection{Implications for Minkowskian spectral reconstruction}
\label{subsec:minkowski-implications}

The Fukuda--Kugo test has a direct bearing on the Minkowskian spectral
problem.  In subcritical coupling the scalar mass function is positive, smooth
and monotonically decreasing in the spacelike domain.  In that regime,
smoothness, monotonicity and positivity-inspired penalties can stabilize
training without distorting the physics.  Near and above the critical region,
however, the same benchmark develops zero crossings and its behavior signals
the breakdown of the assumptions behind a positive Lehmann representation.
A hard positivity constraint then produces a smooth but physically incorrect
solution, one that looks well-behaved precisely because the relevant feature
has been suppressed.
For the dispersive reconstruction, this means that a future spectral PINN
should learn both the scalar dressing and the spectral density,
\begin{equation}
B_\theta(p^2),
\qquad
\rho_{s,\theta}(\omega),
\end{equation}
with the subtracted dispersion relation enforced through the residual
\begin{equation}
\mathcal{R}_{\rm disp}
=
B_\theta(p^2)
-
m(\mu)
-
\int_{m^2}^{\infty}
\dd\omega\,
\frac{
\rho_{s,\theta}(\omega)(p^2 - \mu^2)
}{
(p^2 - \omega + \ii\epsilon)(\omega - \mu^2)
}.
\end{equation}
The positivity of $\rho_{s,\theta}$ should then emerge as an output to be
analyzed.  It can be used as a soft prior in regimes where the Lehmann
representation is expected to hold, but it should not be imposed as a hard
constraint in all regimes.
The dispersive numerical analysis emphasizes three features that are relevant here.
First, the Euclidean Fukuda--Kugo solution and the spectral Minkowskian
solution agree in the weak-coupling regime, up to numerical and
scheme-dependent differences.  Second, the modified unitary equations improve
the comparison with the Euclidean solution by regulating the strong infrared
enhancement of the Lehmann weights.  Third, near couplings of order one the
spectral solution becomes unreliable or ceases to exist, while the Euclidean
equation can still be solved formally.
Our results reproduce the part of this picture that can be isolated cleanly
from the Fukuda--Kugo equation.  The spacelike mass function is smooth and
positive at small $\alpha$, becomes strongly suppressed near
$\alpha_c = \pi/3$, and develops sign changes above the critical region.
The neural reconstruction succeeds only when the output space is flexible
enough to represent those sign changes.  This supports using neural solvers
as diagnostic tools for the analytic regime of the solution, rather than as
black-box interpolators whose assumptions may not hold.

The present calculation does not claim to replace a full solution of 
unitary equations.  It furnishes a controlled neural stress test anchored in
the same quenched rainbow dynamics.  The natural next step is to combine the
neural representation with the unitary-equation residuals, learning the
Lehmann weights and the scalar dressing simultaneously.  That extension would
enable one to test directly whether the neural spectral density remains
positive, becomes sign-indefinite, or fails to support a stable dispersive
reconstruction near the critical regime.

\section{Conclusions and outlook}
\label{sec:conclusions}

We have studied neural reconstructions of DSEs in
Minkowski QED under a quenched, rainbow, Landau-gauge truncation.  The central
reference point is the dispersive momentum-subtraction framework, in which
propagators are reconstructed from Lehmann weights and subtracted dispersion
relations.  This approach makes the analytic assumptions visible: the
existence of a positive Lehmann representation is part of the physical content
of the solution, and its failure is tied to the loss of particle-like
propagation.
One outcome of our present work is largely conceptual.  We have kept separate three
objects that are sometimes conflated: the Euclidean Fukuda--Kugo equation,
unitary equations for the Lehmann weights, and the infrared-regulated
modified unitary equations.  The first is a clean spacelike benchmark for
the quenched rainbow gap equation.  The second defines the genuinely spectral
Minkowskian construction.  The third introduces an infrared cutoff in the
principal-value terms and should be understood as a regulated numerical
variant of the spectral system.  Keeping these distinctions clear is essential
when using neural networks as surrogates or residual solvers.

We solved the Fukuda--Kugo equation directly as a Volterra equation with the
normalization $m(0) = 1$.  The spacelike mass function remains positive
below the critical region and is strongly suppressed as $\alpha$ approaches
\begin{equation}
\alpha_c = \frac{\pi}{3}.
\end{equation}
On the main finite spacelike interval used in the scan, the first observed
zero crossing appears between $\alpha=1.075$ and $\alpha=1.10$.  A refined
scan with increasing ultraviolet endpoint shows that this finite-window
threshold moves toward the analytical value $\alpha_c=\pi/3$.  The apparent
shift is therefore a consequence of using the first zero crossing inside a
finite numerical domain as a practical diagnostic, rather than evidence for a
different critical coupling.  Above the critical region the mass function
develops negative branches and can no longer be described by weak-coupling
intuition.

Our unconstrained neural ansatz accurately reproduces the Fukuda--Kugo solution
from weak coupling through the supercritical regime, including the zero
crossings.  A positive-output ansatz fails once the benchmark develops a
negative branch.  This is not an optimization artifact but the direct
consequence of imposing an output constraint incompatible with the physical
benchmark.  At $\alpha = 2.0$, for instance, the relative error increases
from roughly $10^{-4}$ for the unconstrained ansatz to roughly
$9 \times 10^{-2}$ for the positive-output one.
The lesson for Minkowskian PINNs is worth stating plainly.  Smoothness,
monotonicity and positivity-inspired penalties can be useful stabilizers in the
subcritical regime, where the mass function is smooth, positive and
decreasing.  Near the critical region, those same constraints can become
misleading.  A neural DSE solver should be designed to both reproduce
regular weak-coupling solutions and to detect when the assumptions behind
those solutions are breaking down.  In the present context, spectral
positivity is most useful as a diagnostic of the validity of the Lehmann
representation; imposing it everywhere conflates a physical criterion with an
architectural bias.

Our Fukuda--Kugo neural
reconstruction is a stress test for the architecture and loss design, not a
replacement for a full solution of unitary equations.  A full spectral
PINN should learn, at minimum, both
\begin{equation}
B_\theta(p^2), \qquad \rho_{s,\theta}(\omega),
\end{equation}
while imposing the subtracted dispersion relation as a residual.  In that
formulation, the sign and positivity properties of $\rho_{s,\theta}$ become
outputs to be interpreted, not assumptions to be imposed.

Several extensions follow naturally.  The rainbow truncation should be
replaced by symmetry-preserving vertices such as the Ball--Chiu or
Curtis--Pennington constructions, which would enable the neural solver to test
Ward--Takahashi identities and multiplicative renormalizability more directly.
The photon equation should be unquenched, coupling the fermion and photon
spectral functions in a self-consistent system.  Uncertainty-aware versions of
the method, such as Bayesian PINNs or ensemble-based neural spectral solvers,
could quantify the growing instability as the critical region is approached.
And the same framework extends naturally to Bethe--Salpeter equations \cite{PhysRev.84.1232}, where
Minkowskian analyticity and spectral support are central to the description of
bound states.
This result also fits into our wider line of work connecting continuum field
theory and neural representations.  Together with our Euclidean PINN analysis
of QED DSEs~\cite{Terin2025SciPostCore} and our
gauge-covariant stochastic neural-field framework for stability and
finite-width effects~\cite{Terin2026SciRep}, the present Minkowskian
benchmark features three complementary ingredients: stability diagnostics,
Ward-type constraints, and analytic structure.  These ingredients provide a
natural route toward quantum chromodynamics-inspired systems, scalar Yukawa models,
symmetry-preserving truncations, and full spectral PINNs.

Lastly, the broader point from this work is that neural solvers for continuum quantum field theory
should be constrained by analytic structure, but not overconstrained by
properties that hold only in restricted phases.  In weak coupling, a neural
representation can serve as a compact differentiable surrogate for a smooth DSE
solution.  Near the loss of a Lehmann representation, the same network becomes
more valuable as a diagnostic: it can reveal whether the assumed spectral
structure remains self-consistent, becomes sign-indefinite, or breaks down
altogether.

\acknowledgments

R.C.~Terin thanks Vladimír Šauli for sending a helpful reference.

\bibliographystyle{JHEP}
\bibliography{refs}

\end{document}